\documentclass[a4paper,11pt]{article}

\usepackage{natbib,aas_macros}
\usepackage{amssymb}
\usepackage{xcolor}
\usepackage{fancyhdr}
\usepackage{sectsty}
\usepackage{graphicx}
\usepackage{url}

\usepackage{paralist}
\usepackage{hyperref}
\let\olditem\item
\renewenvironment{thebibliography}[1]{%
  \section*{\refname}
  \let\par\relax
  \renewcommand{\item}[1][]{\olditem[\textbullet]}%
  \inparaenum}{\endinparaenum}

\fancypagestyle{plain}{%
  \fancyhf{}\fancyfoot[R]{Page \thepage}%
}

\begin{document}

\title{
\vspace{-2.5cm} {\em 
{\bf \Large Black-hole activity feedback across vast scales}}}


  \author{{\bf Michal Zaja\v{c}ek} (Masaryk University, Brno; \href{mailto:zajacek@mail.muni.cz}{zajacek@mail.muni.cz}),\\ {\bf Bo\.{z}ena Czerny} (Center for Theoretical Physics PAS, Warsaw),\\ {\bf Rainer Schödel} (IAA-CSIC, Granada), {\bf Norbert Werner} (Masaryk University, Brno),\\ {\bf Vladimír Karas} (Astronomical Institute CAS, Prague) \vspace{0.5cm} \\
 }

\date{}
\maketitle
\vspace{-1.cm}

\begin{abstract}
Black-hole activity feedback is intensively studied on both galaxy-cluster scales and parsec scales. There are, however, many open questions about how the close surroundings of supermassive black holes affect large-scale structure and vice versa. 
\end{abstract}

Both observational and theoretical studies of black-hole activity or active galactic nucleus (AGN) feedback have been ongoing since the first indication of supermassive black holes powering quasar activity in the 1960s. Although several crucial astrophysical questions have been answered in the following decades, a number of open problems remain, in particular how AGN feedback operates over nearly eight orders of magnitude -- from scales of $\sim 10^{-3}\,{\rm pc}$ to the galaxy-cluster scales of a few hundred kiloparsecs. At the beginning of June 2022, about 50 junior as well as senior researchers met in Brno for the post-lockdown edition of the Cologne-Prague-Brno meeting (`\textit{Black-hole activity feedback from the Bondi-radius to galaxy-cluster scales}';  \url{https://cpb2022.physics.muni.cz/}) to try to connect the dots. 

\section*{Feedback on small scales}

Supermassive black holes (SMBHs) are not isolated. On the contrary, they are typically surrounded by the densest stellar systems in galaxies, nuclear star clusters (NSCs), which have comparable sizes to globular clusters but are many times more massive, and, in barred spiral galaxies, nuclear stellar disks, which have scales from a few tens to a few 100 pc (Rainer Schödel, Álvaro Martínez). On the scale of a few 1\,,000 gravitational radii, accretion flows are present that manifest themselves in the form of broad lines, characteristic continuum emission as well as outflows and jets  (Bo\.{z}ena Czerny, Ana Müller, Peter Boorman).

The nearest NSC, and the only one that can be resolved observationally into individual stars (Andreas Eckart, Florian Peissker), is the NSC at the center of the Milky Way (MW NSC) with a half-light radius of $\sim 4$ pc and a stellar mass of $\sim 10^7\,M_{\odot}$. The MW NSC is old ($\sim 80\%$ of the stellar mass formed $\gtrsim 10$ Gyr ago) and possesses a power-law density cusp of stars around Sgr~A*. Past accretion activity of Sgr A* a few million years ago may have affected stars in the surrounding NSC via star-disk and star-jet collisions (Michal Zaja\v{c}ek, Anabella Araudo, Petr Kurfürst), thus stripping away the envelopes of giant stars and rendering them invisible, which may explain the observed dearth of giant stars within about 0.2 pc from Sgr~A*.

More generally, black-hole accretion in AGNs can induce outflows. The model of a failed dusty outflow has been invoked to explain the formation of broad-line regions at a few thousand gravitational radii (Bo\.{z}ena Czerny, Ana Müller), one of the most important observational characteristics of type I AGNs. General relativistic magnetohydrodynamic 2D and 3D models of magnetized accretion flows show that ultrafast outflows and jets can be generated within a few tens to a few hundred gravitational radii (Agnieszka Janiuk, Bestin James). Since SMBHs are surrounded by NSCs, accretion flows can frequently be perturbed by a star or a compact remnant, which can address some of the quasi-periodic accretion or outflow phenomena, including quasiperiodic eruptions (QPEs; Petra Suková).

Several blazars exhibit jet morphological changes, which can be linked to flux density variations and outbursts through changes to the viewing angle. A model of bulk jet precession driven by the Lense-Thirring effect or by a secondary black hole can naturally account for the periodic variation of the Doppler-boosting factor, which changes the underlying jet flux density (Silke Britzen). This model has successfully been applied to address recurring radio outbursts of the `Rosetta-stone' BL Lac source OJ287 or to explain the formation of the ring-like temporal pattern in the radio jet structure of the flat-spectrum radio quasar PKS 1502+106, which is a candidate source of high-energy neutrinos. Exciting prospects for novel observational tests of these models are finally enabled by just-emerging X-ray polarimetry thanks to the Imaging X-ray Polarimetry Explorer (IXPE; Giorgio Matt).           

\section*{Feedback on large scales}

The second part of the conference was dedicated to galaxy clusters, the largest bound structures in the Universe and at the same time the largest structures where AGN feedback operates. Most of the baryonic content in galaxy clusters is in the form of hot, diffuse X-ray emitting gas (with number densities $10^{-4} \lesssim n_{\rm e} \lesssim 10^{-1}\,{\rm cm^{-3}}$ and temperatures $10^6< T_{\rm e} \lesssim10^8\,{\rm K}$) that is approximately in hydrostatic equilibrium (Norbert Werner). Although the hot halo gas is cooling down radiatively mostly through thermal bremsstrahlung, it is prevented from runaway cooling and thus forming stars by AGN feedback in galaxies with halo masses $\gtrsim 10^{12}\,M_{\odot}$, whereas supenovae provide most of the feedback in less massive galaxies \citep{2017NatAs...1E.165H}. Massive, optically bright elliptical galaxies at the centers of galaxy clusters are typically low-luminous AGN that provide radio-mechanical feedback via jets that inflate buoyantly rising cavities detected as depressions in the X-ray surface brightness. A study of 42 optically bright, X-ray emitting giant elliptical galaxies implies a long duty cycle of the radio-mechanical feedback that provides the required heating rate to stabilize the cooling hot atmospheres (Romana Grossov\'a). To correctly analyse the temperature and the metallicity radial profiles of the X-ray emitting gas in galaxy clusters, advances in X-ray spectral analysis are crucial; in particular, the application of spectral deconvolution using the machine learning is useful to assess different spectral models (Carter Rhea).

\section*{Eight orders of magnitude}

Jets emanating from giant elliptical galaxies provide enough radio-mechanical power to inflate radio lobes and drive weak shocks into the intergalactic medium, which balances the radiative cooling. To drive the jet, an AGN needs to be switched on at a moderate level, hence a certain amount of gas needs to be provided for accretion. In galaxy-cluster or large-scale AGN feedback, the theory of `precipitation' is popular \citep{2013MNRAS.432.3401G,2015ApJ...803L..21V}. It is based on thermal instabilities operating in the hot halo gas, thanks to which gas condenses or `precipitates' onto cold clumps or filaments that `rain' onto the central galaxy, and subsequently ignite or maintain AGN feedback. This mechanism has recently been confirmed by the study of the correlation between jet power and Bondi-accretion power. While giant elliptical galaxies that exhibit cooling flows revealed via [NII]$+$H$\alpha$ emission exhibit a significant correlation between jet power and Bondi-accretion power, those without cooling flows lack the same correlation (Tomáš Plšek).  

In comparison, from the point of view of small-scale feedback, accretion onto supermassive black holes is driven by channeling cold gas from larger galactic scales, for example through the bar potential and/or cloud-cloud collisions, which probably caused significant past accretion activity of Sgr~A*.

Of course, both the large-scale and the small-scale effects of the black-hole activity are causally linked. Accretion from the circumgalactic medium, in the form of precipitation, for example, is a primary source of material for both star-formation as well as accretion and AGN feedback. However, the AGN feeding and feedback act on different timescales, which makes the comprehensive statistical analysis challenging. 

\begin{figure}
    \centering
    \includegraphics[width=\textwidth]{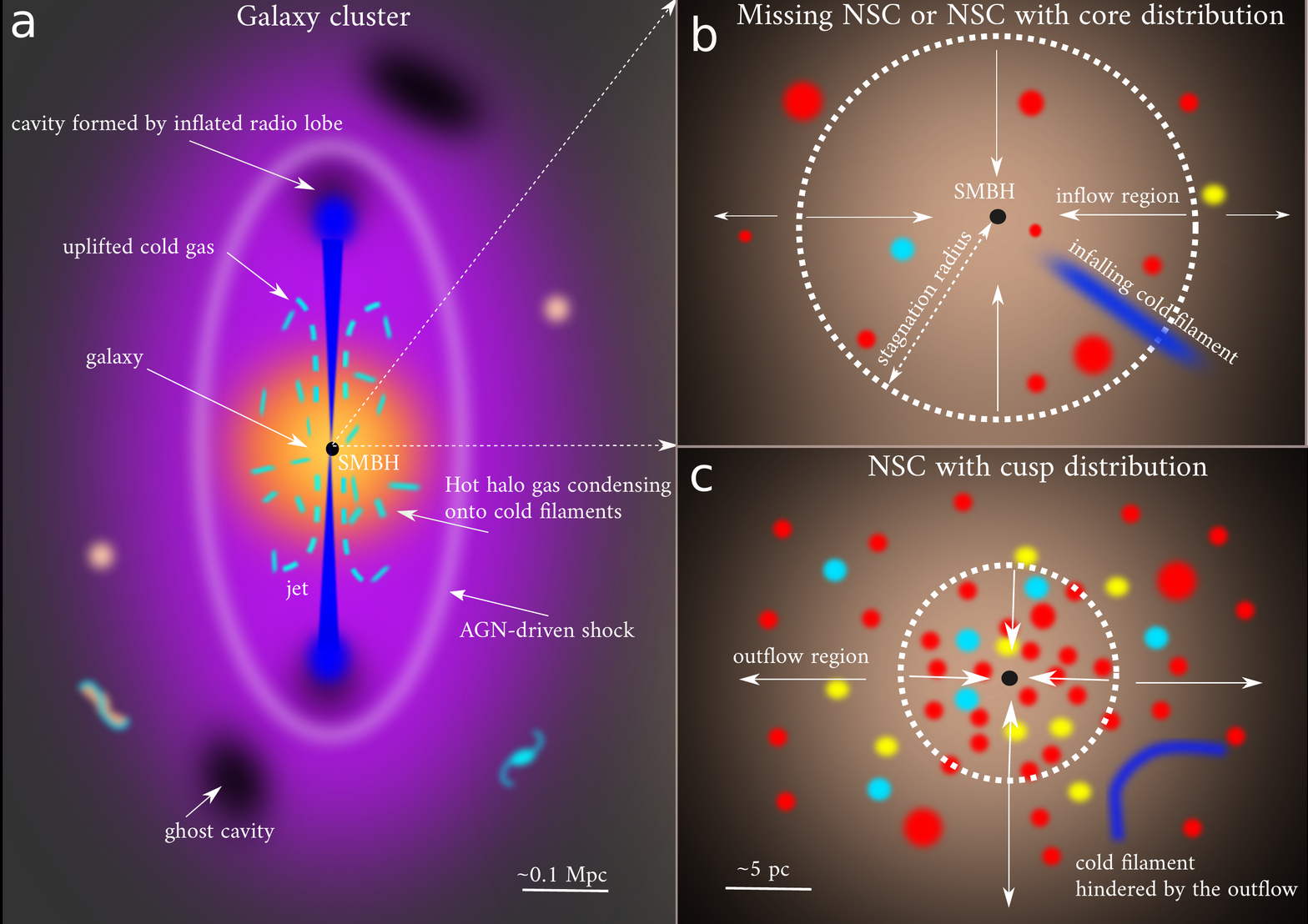}
    \caption{{\bf Mutual relation between large-scale AGN feeding and feedback, and the processes taking place within the sphere of influence of an SMBH.} {\bf a,} The surroundings typical for galaxies found at the centers of galaxy clusters. Hot halo gas is prone to thermal instability, which leads to the condensation of the hot plasma onto the cold clumps and filaments. This cold gas then `precipitates' onto the galaxy where it powers the AGN radio-mechanical feedback in the form of the jet that inflates radio lobes that push away the hot plasma. This effect can be detected in the form of depressions in the X-ray surface brightness or cavities. Radio-mechanical feedback that stabilizes the hot halo atmosphere originates in galactic nuclei hosting typically at least one SMBH. The two yellow disks and blue spirals represent smaller galaxies within the galaxy cluster. {\bf b, c,} When SMBHs are surrounded by nuclear star clusters, the latter may moderate the accretion onto the black hole. The cluster stellar density profile -- a flat core (or a missing NSC in the extreme limit) ({\bf b}) or a cusp ({\bf c}) -- influences the classical Bondi accretion due to the presence of wind-blowing stars. The radial flow velocity goes through zero at the stagnation radius, which divides the inflow and the outflow regions. While for the cusp-like NSC ({\bf c}) the stagnation radius is comparable to the Bondi radius, for the core-like NSC ({\bf b}) it is about twice as large. This clearly directly affects the volume, from which the SMBH can accrete colder gas. This in turn influences the accretion rate as well as the jet power that acts further on kpc to Mpc scales.}
    \label{fig_AGN_feedback}
\end{figure}

Parsec-scale nuclear star clusters can in principle mix up the cards. In the stationary 1D models of the stellar-wind feeding in the NSCs \citep{2015MNRAS.453..775G,2018MNRAS.479.4778Y}, the radial flow velocity goes through zero at the so-called stagnation radius, which separates inflow and outflow regions. In this sense, NSC can `isolate' the black hole, which is then mostly fuelled -- at a very low level --  by the stellar-wind material within the inflow zone while any external cold material is hindered by the collective NSC outflow. The final accretion and feedback state of the galaxy is then modulated by the NSC presence, its stellar brightness profile (core or cusp), and the stellar-wind properties of late-type or early-type stars. This is illustrated in Fig.~\ref{fig_AGN_feedback}, where we depict the connection between the large-scale feeding and AGN feedback on megaparsec scales, and the processes operating on the scale of a few parsecs within an NSC.     

\section*{Ernst Mach and Brno}

The combination of detailed hydrodynamical modelling as well as high-resolution X-ray and optical data is required to understand the complex interplay between the large-scale cold gas inflow and the small-scale distribution of matter within the gravitational sphere of influence of the SMBH. The connection between large-scale and small-scale properties of galaxies was the central theme of the conference. This is related to \textit{Mach's principle} stating that the local state of inertia is determined by the large-scale distribution of matter
\citep{1995mpfn.conf.....B}. Like the main theme of feeding and feedback near cosmic black holes, Mach's principle has a continued impact on all scales of the gravitational Universe. Indeed, in general relativity, the effects of frame dragging influence particles along with fields (Vladimír Karas).

It is, therefore, extraordinarily fitting that the meeting happened to take place in Brno, close to Ernst Mach's birth place of Chrlice, where he was born in 1838 (Ji\v{r}í Dušek). Moreover, the meeting held the award ceremony for the Ernst Mach Medal, an honorary award of the Czech Academy of Sciences (Eva Zažímalová). This year, the recipient was Andreas Eckart, in recognition of his life-long contribution to physical sciences, in particular for his important contribution to infrared studies of the Galactic Centre. Eckart has contributed to both observational techniques as well as to the theory of astronomical black holes, additionally using the framework of philosophical discussions besides standard approaches \citep{2017FoPh...47..553E}. He contributed to the development of decisive infrared instrumentation, in particular SHARP and GRAVITY, for the European Southern Observatory telescopes. In addition, he conducted the first high-resolution near-infrared imaging observations of the vicinity of Sgr~A*, which resulted in the detection of fast-moving stars around Sgr~A*, known as S-stars. These are now routinely used for the tests of general relativity, developed by Albert Einstein, who was in turn inspired by Mach's principle.



\section*{Acknowledgements}
The organizers of the Cologne-Prague-Brno meeting thank the Grant Agency of the Czech Republic for the financial support via the EXPRO grant no. 21-13491X ``Exploring the Hot Universe and Understanding Cosmic Feedback'' and the Czech-German collaboration project no. 19-01137J ``Largest Black Holes in the Sky: Origin and Evolution of Horizon-Scale Structure''. We also thank the Brno Observatory and Planetarium (J. Dušek, Z. Kuljovská, M. Mesarč) for providing the venue and the social programme. We are grateful to the volunteers from the Department of Theoretical Physics and Astrophysics of the Faculty of Science, Masaryk University (V. Glos, J.-P. Breuer, K. Protušová, N. Husáriková, L. Santarová, F. Münz) for the technical and the administrative help during the conference.

\end{document}